\documentstyle[preprint,prb,aps]{revtex}
\begin{document}
\draft
\title{Plasmon-pole approximation for
semiconductor quantum wire electrons}
\author{S. Das Sarma, E.H. Hwang, and Lian Zheng} 
\address{Department of Physics,
University of Maryland, College Park, Maryland 20742-4111}
\date{\today}
\maketitle
\begin{abstract}
We develop the plasmon-pole approximation for an interacting electron 
gas confined in a semiconductor quantum wire. We argue that
the plasmon-pole approximation becomes a more accurate approach
in quantum wire systems than in higher dimensional systems 
because of severe phase-space 
restrictions on particle-hole excitations
in one dimension. 
As examples, we use the plasmon-pole approximation to calculate
the electron self-energy due to the Coulomb interaction
and the hot-electron energy relaxation rate
due to LO-phonon emission 
in GaAs quantum wires. We  find that the plasmon-pole
approximation works extremely well
as compared with more complete many-body calculations.
\end{abstract}
\narrowtext
\section{introduction}
\label{one}
Recently, there has been an increasing interest \cite{rev1}
in semiconductor quantum wire structures, where the motion of
electrons is essentially restricted
to be one dimensional.
Technological progress has made it possible to fabricate \cite{gon1}
high quality quantum wires
where only the lowest subband is populated by electrons,
so that a truly one-dimensional interacting  
electron gas is realized.
Much in the same way as quantum well structures have generated 
tremendous activities in pure and applied research on two-dimensional 
electron systems, quantum wire structures have created the potential  
for new device applications \cite{rev1,apl1,apl2}
and the opportunity to carry out experimental study on 
one-dimensional Fermi systems, where many theoretical predictions\cite{lut} 
can be tested.
Because of the low dimensionality, properties of  
a quantum wire 
are very sensitive to electron-electron interaction effects \cite{lut,hd1}.
Many experimentally relevant quantities 
need to be calculated by taking into account 
many-body interaction induced exchange-correlation effects. 
The standard perturbation theories,
which have been developed for higher dimensional electron systems,
have been applied \cite{hd1,li1} to quantum wire systems,
and good agreement with experiments \cite{gon1,exp1}
are generally obtained.
In this article, we discuss the application of 
another well known many-body approach, the plasmon-pole approximation,
\cite{ppa1,ppa2} to quantum wire systems.  The motivation 
for this work is the observation 
that the collective plasmon excitation plays a more prominent role 
in  a one-dimensional electron system
compared with its higher-dimensional counterparts because 
single-particle electron-hole excitation continuum 
is severely restricted in one dimension due to 
energy-momentum conservation. 
Thus, the plasmon-pole approximation,
besides having the obvious benefit of great simplicity,
may work well for one-dimensional 
quantum wires in calculating exchange-correlation effects.
To illustrate this point, we calculate the electron self-energy correction 
due to electron-electron Coulomb interaction 
and the electron energy relaxation rate due to electron LO-phonon 
Fr\"ohlich interaction
and compare the plasmon-pole approximation results 
with the corresponding full 
many-body calculations using the 
random-phase approximation (RPA).

The plasmon-pole approximation 
has been extensively employed \cite{ppa1,ppa2,ppa3,ppa4}
in calculating the electron self-energies of three- and two-dimensional 
systems. The results obtained from these calculations 
are in good semiquantitative agreement with the results of more 
sophisticated treatments---namely the full RPA calculations,
and with experimental results. 
A many-body interacting electron system has  
both collective plasmon excitations
and single-particle electron-hole excitations\cite{man1}. 
The plasmon-pole approximation simplifies the many-body excitation spectrum 
by ignoring 
the particle-hole excitations and assigning the whole spectral weight,
which is dictated by the f-sum rule \cite{man1}, 
to an effective collective plasmon excitation,
which is assumed to be a real pole of the response function.
This is, in general, a crude approximation
for the actual dynamical response of the electron system,  
except in the long wavelength limit
where the plasmon excitation exhausts all the spectral weight 
in a uniform system \cite{man1} by virtue of particle conservation.
It is well known that in a one-dimensional system, 
the collective plasmon excitation plays a more prominent role
because of phase space restriction
on particle-hole excitations. 
In fact, the long wavelength RPA plasmon dispersion is 
exact in one-dimensional electron liquids up to second order in
wavevector in contrast to higher-dimensional systems \cite{hwang}.
It is therefore worthwhile to explore the possibility
that the plasmon-pole approximation may actually work
better in quantum wire systems  
than in higher dimensional systems.
Our work is motivated by this purpose. 
We find that the one-dimensional phase-space 
restriction on the particle-hole 
excitations indeed increases the spectral weight
of the plasmon excitation over a wide
range of wavevectors
in one-dimensional systems 
under conditions which are typical in GaAs-based quantum wire samples,
and the plasmon-pole approximation indeed works extremely well in
calculating a variety of quantities in GaAs-based quantum wire structures. 
We specifically apply the plasmon-pole
approximation to two different problems: one involves the electron-electron 
Coulomb interaction and the other involves the polar electron LO-phonon
Fr\"ohlich interaction. In the first case, 
we calculate quasiparticle properties of the 
interacting electrons
by taking into account the 
Coulomb interaction effects through the plasmon-pole approximation.  
In the second case, 
we calculate the hot electron energy-loss rate
via LO-phonon emission. 
We show that the plasmon-pole approximation works well
in both cases by giving results 
which are in good agreement with the corresponding RPA results,
which are much more difficult computationally.

In Fig. \ref{f11}, we show the elementary
excitation spectrum of a one-dimensional electron system
calculated within the RPA,
where the particle-hole excitations are confined within
the phase-space
surrounded by the dotted-line $ABCDE$, and 
the plasmon excitation
is represented by the solid-line.
In one dimension, 
the RPA plasmon dispersion has a simple analytical expression \cite{li1}  
(we set $\hbar=1$ throughout this paper)
\begin{equation}
\omega_q={A(q)E_+(q)-E_-(q)\over A(q)-1},
\label{equ:w1}
\end{equation}
where $E_{\pm}(q)=q^2/2m\pm k_Fq/m$ with $m$ as electron mass
and $k_F$ the electron Fermi wavevector,
$A(q)=\exp[q/\pi V_c(q)]$ with $V_c(q)$ as the electron-electron 
Coulomb interaction potential \cite{1d1p}
of a quantum wire of finite lateral width. Unlike plasmon modes 
in higher dimensions, the RPA plasmon excitation in a quantum wire
exists ({\it i.e.} is undamped) for all wavevectors $0\leq q<\infty$. 

The most characteristic feature of the one-dimensional spectrum
is that particle-hole excitations are prohibited
from a large portion of the low energy phase space, the region below
the dotted-line $BCD$ in Fig. \ref{f11}.
This restriction, which arises from
the momentum-energy conservation,
increases the dominance of plasmon excitations
in a quantum wire compared with higher-dimensional systems.
The situation is totally different in
higher dimensional systems \cite{man1}, 
where particle-hole
excitations are allowed in the whole phase space between
the dotted-lines $AB$ and $DE$ in Fig. \ref{f11}.
For a quantitative measure of the relative importance 
of the plasmon excitation,
we evaluate its oscillator strength within the RPA
\begin{equation}
F(q)=-{2m\over\pi nq^2}\int_0^\infty\omega S_{PL}(q,\omega)d\omega, 
\label{equ:os1}
\end{equation}
where $n$ is the average density of the electron gas, and
the plasmon spectral weight is defined by
(with the plasmon dispersion $\omega_q$ given 
by Eq. (\ref{equ:w1}))
$$S_{PL}(q,\omega)=-{\pi\over V_c(q)}{1\over\left|{\partial\over\partial\omega}
{\rm Re}[\epsilon(q,\omega)]\right|}\delta[\omega-\omega_q].$$
In Fig. \ref{f12}, we compare the oscillator strengths of plasmon excitation 
of a one-dimensional quantum wire with that of a 
two-dimensional quantum well (both obtained within the RPA).
The plasmon oscillator strength in the quantum well
drops quickly to zero at a critical wavevector, beyond which
an undamped well-defined plasmon mode does not exist. 
The oscillator strength of the plasmon excitation 
in a quantum wire,
on the other hand,  extends well into the range of large wavevectors,
decreasing slowly with increasing wavevector. 
Note that the $q\rightarrow0$ behavior of these curves, {\it i.e.} 
$F(q)=1$ for $q\rightarrow0$, is just a manifestation of the f-sum rule.
The interesting point is that $F(q)\sim1$ in one dimension even for \linebreak
$q>k_F$.
It is seen clearly that the plasmon dominance 
of the spectral weight is
significantly increased in quantum wire systems.
Since the plasmon-pole approximation assumes 
that the excitation spectrum consists
of no particle-hole excitations, but
solely of a collective mode which exists for all values of wavevectors
and possesses unit oscillator strength,
it is easy to understand why the plasmon-pole approximation
may work well in a quantum wire system.
The density-density response function 
of a quantum wire electron system 
in the plasmon-pole approximation is given as
\begin{equation}
\chi_{PP}(q,\omega)={{n\over m}q^2\over\omega^2-\omega_q^2}.
\label{equ:x1}
\end{equation}
By construction, $\chi_{PP}(q,\omega)$ in the above expression 
satisfies the f-sum rule and
the static Kramers-Kronig relation \cite{man1}. 
Eq. (\ref{equ:x1}) is our plasmon-pole approximation model
\cite{ppa1,ppa2,ppa3,ppa4}, which we use to calculate 
quantum wire many-body electronic properties.

In Sec. \ref{self} and \ref{err}, we apply the plasmon-pole approximation 
to the calculations of electron self-energy due to
Coulomb interaction and electron energy relaxation rate due to 
LO-phonon emission, respectively. 
A short summary in Sec. \ref{sum} concludes our paper.

\section{self-energy and spectral function}
\label{self} 
The one-dimensional (1D) 
self-energy within the leading order GW approximation\cite{hedin}
neglecting vertex correction at $T = 0$ is given by 
\begin{equation}
\Sigma(k,\omega) = i\int \frac{dqd\omega'}{(2\pi)^2} W(q,\omega)
G_0(k-q,\omega-\omega'), 
\end{equation}
where $G_0(k,\omega)$ is the Green's function for the noninteracting
electron gas 
\begin{equation}
G_0(k,\omega) = \frac{\theta(|k|-k_F)}{\omega - \xi(k) + i 0^+} +
\frac{\theta(k_F - |k|)}{\omega - \xi(k) - i 0^+},
\end{equation}
with $\xi(k) = k^2/2m - \mu$, ($\mu$=chemical potential) and $W(q,\omega)$
is the dynamically screened Coulomb interaction, which is given by
\begin{equation}
W(q,\omega) = \frac{V_c(q)}{\epsilon(q,\omega)}.
\end{equation}
Here $V_c(q)$ is the bare Coulomb interaction, which is
logarithmically divergent in the 1D wavevector space. Thus, we use
the more realistic finite width quantum wire model whose  fully
approximated matrix element can be found in the literature\cite{1d1p}.  
$\epsilon(q,\omega)$ is the dielectric function, which describes 
the dynamical 
screening properties of the electron gas. The dynamically 
screened interaction
$W(q,\omega)$ 
can be separated into an unscreened term which gives rise to the
exchange part of the self-energy and another term which gives rise to the
correlation part of the self-energy and involves coupling to
density fluctuations
\begin{equation}
W(q,\omega) = V_c(q) + V_c(q) \left [ \frac{1}{\epsilon(q,\omega)} - 1
\right ].
\end{equation}
The imaginary part of the second term is nonzero within the
electron-hole continuum and along the plasmon dispersion line in
the RPA. In the plasmon-pole
approximation (PPA)\cite{ppa1,ppa2,ppa3,ppa4} the second term is
replaced by a coupling to the effective plasmon
mode as described \linebreak in Sec. \ref{one}
\begin{equation}
{\rm Im} \left [ \frac{1}{\epsilon(q,\omega)} - 1 \right ] =
-\frac{\pi}{2} \frac{\omega_{0}^2}{\omega_q} \delta(\omega -
\omega_q),
\label{epsilon}
\end{equation}
where the strength 
$\omega_{0}^2 = \frac{n}{m} V_C(q) q^2 $ is determined by the
requirement that Eq. (\ref{epsilon})  satisfies the $f$-sum rule,
and $\omega_q$ is the 1D plasmon
dispersion which is exactly known within RPA \cite{li1}
(See Eq. (\ref{equ:w1})). 
Unlike in 2D and 3D, where the exact analytic RPA plasmon
dispersion is unknown so that the static RPA dielectric function
$\epsilon(q,\omega=0)$ is used
in obtaining the effective plasmon frequency $\omega_q$, 
we use the analytically known 1D RPA plasmon
dispersion given in Eq. (\ref{equ:w1}). Note that any 
attempt to use the static RPA (similar to what is done in 2D and 3D PPA)
in 1D PPA is not only unnecessary (because the 1D RPA plasmon 
dispersion is known analytically), but also incorrect because 
the 1D static RPA
dielectric function has logarithmic zero temperature 
singularities due to a divergence at $q=2k_F$.  
Using the Kramer-Kronig relation we have
\begin{equation}
\frac{1}{\epsilon(q,\omega)} -1 = \frac{\omega_0^2}{\omega^2 -
\omega_q^2 + i\delta}.
\label{eq9}
\end{equation}
Within the PPA the self-energy can now be separated into a frequency
independent exchange term
and a correlation term
\begin{equation}
\Sigma(k,\omega) = \Sigma_{\rm ex}(k,\omega) + \Sigma_{\rm
cor}(k,\omega),
\end{equation}
where
\begin{equation}
\Sigma_{\rm ex}(k,\omega) = i \int \frac{dqd\omega'}{(2\pi)^2} V_c(q)
G_0(k+q, \omega + \omega'),
\end{equation}
and
\begin{equation}
\Sigma_{\rm cor}(k,\omega) = i \int \frac{dqd\omega'}{(2\pi)^2} V_c(q) \left [
\frac{1}{\epsilon(q,\omega')} - 1 \right ] G_0(k+q,\omega + \omega').
\label{eq12}
\end{equation}
The exchange energy $\Sigma_{\rm ex}(k,\omega)$ 
as well as the correlation energy 
$\Sigma_{cor}$ within the full RPA theory
has been calculated earlier by Hu and Das
Sarma\cite {hd1}. 
Using Eq. (\ref{eq9}) in Eq. (\ref{eq12})
and performing a frequency integration, the
correlation 
part becomes
\begin{equation}
\Sigma_{\rm cor} = \int \frac{dq}{2\pi} \frac{V_c(q)\omega_{\rm
p}^2}{2\omega_q} \left [ \frac{\theta(k_F -|k+q|)}{\omega + \omega_q -
\xi_{k+q} -i\delta} + \frac{\theta(|k+q|-k_F)}{\omega - \omega_q -
\xi_{k+q} +i\delta} \right ].
\end{equation}
The real and the imaginary
parts of the $\Sigma_{\rm cor}$ are given by
\begin{equation}
{\rm Re}\Sigma_{\rm cor}(k,\omega) = {\rm P}\int\frac{dq}{2\pi}g(q) \left
[\frac{\theta(k_F -|k+q|)}{\omega + \omega_q - \xi_{k+q}} +
\frac{\theta(|k+q| + k_F)}{\omega - \omega_q - \xi_{k+q}} \right ],
\end{equation}
and
\begin{eqnarray}
{\rm Im} \Sigma_{\rm cor}(k,\omega) & & = \pi \int\frac{dq}{2\pi}g(q)
\nonumber \\
& &\times \left [ \theta(k_F -|k+q|)\delta(w + w_q -\xi_{k+q}) -
\theta(|k+q|-k_F)\delta(\omega - \omega_q -\xi_{k+q}) \right ],
\label{eq15}
\end{eqnarray}
where $g(q) = V_c(q) \omega_{0}^2/(2\omega_q)$ and P$\int$ indicates
the principle value integral.
From the restrictions on the integration region
arising from various $\theta$ and $\delta$ functions, we
see that ${\rm Im}\Sigma_{\rm cor}$ is nonzero only for
\begin{eqnarray}
\begin{array}{llll}
\omega_q(k_F-k) > \omega & {\rm and} &-\omega_q(k_F + k) < \omega <\xi_k &
\mbox{if $k \leq k_F$}, \\ 
\xi_k > \omega&{\rm and} &-\omega_q(k_F + k ) < \omega <-\omega_q(k-k_F)&
\mbox{if $k > k_F$}.
\end{array}
\label{eq16}
\end{eqnarray}
Carrying out the integral over $q$, one obtains
the imaginary part of the $\Sigma_{\rm cor}$
\begin{equation}
{\rm Im}\Sigma_{\rm cor}(k,\omega) = \frac{1}{2}  \sum_i \left [
\frac{g(q_{+,i}) \theta(k_F - |k+q_{+,i}|)}{\left |
d\Omega_+(q_{+,i})/dq \right |} +
\frac{g(q_{-,i}) \theta(|k+q_{-,i}|-k_F)}{\left |
d\Omega_-(q_{-,i})/dq \right |} \right ],
\label{eq17}
\end{equation}
where $\Omega_{\pm}(q) = \omega \pm
\omega_q - \xi_{k+q}$ and $q_{\pm,i}$ are zeros of $\Omega_{\pm}(q)$.
From Eqs. (\ref{eq15}) and (\ref{eq16}) we know that
Im$[\Sigma(k,\omega)]$ as a 
function of $\omega$ has finite
discontinuities at $\omega = \pm \omega_q(k + k_F)$, whose magnitude
can be calculated from Eq. (\ref{eq17}). For example, we have the magnitude
$g(k_F)/[\partial \omega_q(k_F)/\partial q \pm k_F/m]$ at $\omega =
\pm \omega_q(k_F)$ for $k=0$ and
$(1/2)g(2k_F)/[\partial\omega_q(2k_F)/\partial q \pm k_F/m]$ at
$\omega = \pm \omega_q(2k_F)$ for $k=k_F$. (See the numerically calculated
values in Fig. 3(a) and (b).)  
A finite discontinuity in
Im$[\Sigma]$ gives rise to a logarithmic singularity in Re$[\Sigma]$,
which can be verified using the Kramers-Kronig relation. (See
Fig. 3(a) and (b).)

In order to determine quasiparticle excitation energies one must solve
the Dyson equation\cite{hedin} which is given by
\begin{equation}
\omega + \mu = \xi(k) + \Sigma (k,\omega),
\end{equation}
where $\mu$ is the chemical potential of the interacting electron
gas, which is determined by setting $k=k_F$ and
$\omega=0$ in the above equation. Once the self-energy
$\Sigma(k,\omega)$ is known the 
single particle spectral function $A(k,\omega)$ is readily
calculated. $A(k,\omega)$ contains important dynamical 
information about the
system and is given by
\begin{equation}
A(k,\omega) = \frac{2|{\rm Im}\Sigma(k,\omega)|}{ \left [ \omega -
\xi(k) - {\rm 
Re}\Sigma(k,\omega)\right ]^2 + \left [ {\rm Im}\Sigma(k,\omega)
\right ]^2}.
\end{equation}
It satisfies the sum rule
\begin{equation}
\int^{\infty}_{\infty}\frac{d\omega}{2\pi}A(k,\omega) =1,
\end{equation}
which we verify to be satisfied within less 
than a percent in our numerical
calculations.

Fig. 3 shows the calculated self-energies and spectral functions as a
function of frequency $\omega$ for $k=0$ (band edge) and 
$k=k_F$ (Fermi energy). The complete
RPA results\cite{hd1} (thin lines) are also shown
for comparison with our PPA results. From the figures we 
can see that our PPA
results are almost identical to the full RPA results\cite{hd1}. 
In both calculation the parameters corresponding to 
GaAs are used: $m=0.07m_{\rm e}$ ($m_{\rm e}$ is the free electron mass),
$\epsilon_0 = 12.9$, $\epsilon_{\infty}=10.9$, and $\omega_{\rm LO} =
36.8$\,meV.  The well width of $a=100 \AA$ and the 1D electron density 
of $n=0.56 \times 10^6 {\rm cm}^{-1}$, which corresponds to a
Fermi energy $E_F\approx 4.4$meV and a dimensionless density parameter
$r_{\rm s}=4me^2/\pi k_F \epsilon_0 = 1.4$ with $k_F = \pi n/2$, are
used in both calculation. 
In Figs. 3(a) and (b) the straight lines are given by $\omega - \xi(k)
-\mu$, and their intersections with Re$[\Sigma]$ indicate the solutions
to Dyson's equation and correspond to quasiparticle peaks.
In the spectral function for $k=0$, we find two 
undamped quasiparticle peaks.
The strength ($2\pi \times 0.37$) of the regular
quasiparticle (the first peak near $\omega=0$) within PPA is slightly
higher than the corresponding RPA
result ($2\pi \times 0.33$). The strength ($2\pi \times
0.31$) of the second peak, the so-called plasmaron peak, 
is nearly the same in the PPA as that in the RPA ($2 \pi \times
0.32$). The low energy incoherent spectrum ($E_F < \omega < \omega_q(k_F)$)
arising from the electron-hole 
continuum within RPA is transferred to the quasiparticle spectrum
in the PPA, making the quasiparticle spectral weight 
slightly higher in the PPA than in the RPA.
For $k=k_F$ we find that the quasiparticle-like peak at $\omega=0$ is
not a strict $\delta$-function peak, which means that the system
within PPA has no true long-lived quasiparticles. This result
is qualitatively the
same as the RPA result,
implying that there can be no true quasiparticles in one dimension.
As $\omega \rightarrow 0$, the dominant
contribution to Im$[\Sigma(k_F,\omega)]$ within RPA comes from the plasmon
excitation \cite{hd1}. Therefore, the behavior of the spectral function
near $\omega=0$ for $k=k_F$ shows exactly the same 
behavior for both the PPA and the RPA.
 
In 2D \cite{ppa4} and 3D
\cite{ppa1,ppa2}, the quantitative differences between the results of
plasmon-pole approximation and the random-phase approximation 
are comparable, and are considerably larger than what we find
in our 1D calculations. 
In our 1D calculation, the agreement between the RPA and PPA self-energies 
is almost perfect.
Since the 1D electron-hole continuum is strongly suppressed by
the severe phase restriction due to energy-momentum conservation, the 1D
plasmon is the dominant excitation which contributes
to the electron self-energy, with the contribution from 
single-particle excitations being essentially negligibly small.
In this paper we provide an easy method for calculating the effects of
correlation on the single particle self-energy
in one dimension. 
It should be fairly straightforward to extend the PPA 
self-energy calculation to more complicated experimentally relevant
situations, such as finite temperatures and multisubband 
occupancies, with reasonable confidence of obtaining quantitatively
accurate results. This is the main significance of our work.

\section{energy relaxation in a quantum wire}
\label{err}
In this section, 
we apply the plasmon-pole approximation to a coupled  electron-LO-phonon 
system in a quantum wire and
study hot electron energy relaxation \cite{ja1} through 
LO-phonon emission.
Although this topic is of great importance by itself \cite{ja1},
the  present purpose is to use it as an example to show
the simplicity and the reasonable quantitative accuracy of the plasmon-pole 
approximation in calculating quantum wire electronic properties.
We, therefore, refrain from discussing in details the hot
electron relaxation phenomena \cite{ja2}.

When excess energy is supplied to an electron gas, the electrons 
go out of equilibrium with the underlying lattice, with the  
electron gas attaining an effective electron temperature $T$ higher than 
the embedding lattice temperature $T_L$.
Such a hot electron gas loses energy to its surrounding in order to 
return to equilibrium with the lattice. In polar semiconductors such as GaAs, 
the most efficient energy relaxation process,
except at very low electron temperatures, is through 
LO-phonon emission.
When reabsorption of the emitted LO-phonons is ignored,
the hot electron energy loss rate 
at zero lattice temperature is given by \cite{ja2} 
(we take $T_L=0$ throughout, our results should be valid
for low values of $T_L$):
\begin{equation}
P=\sum_q\int_{-\infty}^\infty
{d\omega\over\pi}\omega n_T(\omega)
|M_q|^2{\rm Im}\chi^{\rm ret}(q,\omega){\rm Im}D^{\rm ret}(q,\omega),
\label{equ:p1}
\end{equation}
where $n_T(\omega)$ is the Bose distribution factor at 
electron temperature $T$, and $|M_q|^2$ is the
Fr\"{o}hlich coupling matrix \cite{li1}. 
The phonon propagator in Eq. (\ref{equ:p1}) is
\begin{equation}
D(q,\omega)={2\omega_{\rm LO}\over\omega^2-\omega^2_{\rm LO}
-2\omega_{\rm LO}|M_q|^2\chi(q,\omega)}.
\label{equ:d1}
\end{equation}
The last term in the denominator is the phonon self-energy
correction due to many-body electron-phonon coupling,
which broadens the phonon spectral function.
The phonon mode couples to the plasmon excitation 
(the so-called plasmon-phonon coupling)
as well as to particle-hole excitations,
so that the renormalized phonon spectrum 
may be characterized as containing hybridized
phonon-like and plasmon-like modes, and quasiparticle-like modes. 
The phonon-like mode has large spectral weight and high energy
($\sim\omega_{LO}$), while the other modes have small spectral weights
but have arbitrarily low energies. 
At high electron temperatures ($k_B T\sim\omega_{LO}$),
energy relaxation through emission of the phonon-like mode
dominates because of its large spectral weight, 
while at low temperatures ($k_B T\ll\omega_{LO}$),
energy relaxation though emission of the plasmon- and 
quasiparticle-like modes dominates because of their low energies.
The existence of the low energy modes enhances the energy loss rate at 
low temperatures since emission of
bare phonon mode with frequency $\omega_{LO}$
is effectively frozen out when $k_B T\ll\omega_{LO}$.
Our present objective is to compare the energy loss rates among 
these three cases: no many-body phonon-electron coupling, involving 
only the bare phonon mode; phonon-electron coupling 
in the plasmon-pole approximation, involving only the hybridized 
plasmon- and phonon-like modes, but no quasiparticle-like modes;
and the phonon-electron coupling in the full RPA, involving all the modes.
One can see that these are three increasingly sophisticated approximations
to the phonon self-energy correction in \linebreak Eq. (\ref{equ:d1})
with the 
phonon self-energy correction completely
neglected in the bare phonon case.

With the phonon self-energy ignored, Im$D(q,\omega)$
becomes a single $\delta$-function at $\omega=\omega_{LO}$. 
Eq. (\ref{equ:p1}) then gives the energy loss rate as
\begin{equation}
P_0=\omega_{\rm LO}n_T(\omega_{\rm LO})\sum_q
(-2)|M_q|^2\chi(q,\omega_{\rm LO}).
\label{equ:po1}
\end{equation}
The characteristic of the bare phonon result is an approximate
exponential temperature dependence $P_0\propto{\rm exp}(-\omega
_{\rm LO}/k_BT)$, which comes from the 
Bose factor $n_T(\omega_{\rm LO})$.
With the plasmon-pole approximation $\chi_{PP}$,
Im$D$ becomes a pair of
$\delta$-functions at $\omega=\omega_{\pm}$, 
the frequencies of the hybridized plasmon-phonon
modes \cite{ja2}.
The energy relaxation rate is then given as
\begin{equation}
P_{\rm PP}=P_++P_-,
\label{equ:tt1}
\end{equation}
with 
\begin{equation}
P_{\pm}=\sum_q\omega_\pm n_T(\omega_\pm)
{\omega_{\rm LO}|\omega_\pm^2-\omega_P^2|\over
\omega_\pm(\omega_+^2-\omega_-^2)}|M_q|^2
(-2){\rm Im}\chi(q,\omega_\pm),
\label{equ:tt2}
\end{equation}
where $P_{\pm}$ refer respectively to energy loss via upper (lower)
hybrid plasmon-phonon modes.
It should be noticed that the above expression is formally 
as simple as the corresponding bare phonon result 
given in Eq. (\ref{equ:po1}), both involving a wavevector integral. 

The energy loss rates with no phonon renormalization, with
phonon renormalization in the plasmon-pole approximation, and
with phonon renormalization in the full RPA, 
are shown in Fig. \ref{f31}.
Two things need to be emphasized. The first is that the 
phonon renormalization enhances 
the energy loss rate by orders of magnitude
at low temperatures, although 
its effect is negligible at high temperatures.
The second is that the plasmon-pole approximation
gives an excellent description of the 
energy loss process, 
in the sense that its result agrees very well with
the full RPA result.
This example shows again that the plasmon-pole approximation
can work remarkably well in a quantum wire system
because of the increased dominance of plasmon excitation 
in one dimension.
\section{summary}
\label{sum}
The plasmon-pole approximation has been widely employed in three-
and two-dimensional many-body electron systems. 
In this work, we discuss two specific 
applications of the plasmon-pole approximation 
to 1D electrons in a quantum wire structure.  Our results suggest 
that the plasmon-pole approximation can work 
exceptionally well in calculating 
electronic many-body properties in a semiconductor
quantum wire structure because of 
the severe phase space restriction on single-particle electron-hole
excitations in one-dimensional systems. 
We apply the plasmon-pole approximation to calculations
of electron self-energy due to Coulomb interactions
and hot electron energy relaxation rate via LO-phonon emission,
and find that our calculated PPA results agree extremely well
with the results of the full RPA calculations.
The agreement of the PPA results with the full RPA results
is substantially better (in fact, essentially exact) in 1D than in 
the corresponding 2D and 3D systems. Our results
should influence future electronic calculations in semiconductor 
quantum wires where calculations may now safely ignore
the full complications of the RPA and adapt the simple, intuitively 
appealing and quantitatively accurate PPA.
\acknowledgments
This work is supported by the U.S.-ONR and the
U.S.-ARO.

\begin{figure}
\caption{Excitation spectrum of a one-dimensional electron gas
in the RPA. 
Particle-hole excitations  are confined within the phase space
surrounded by the dotted-line $ABCDE$. 
Plasmon excitation is represented by the solid-line.
}
\label{f11}
\end{figure}

\begin{figure}
\caption{Calculated RPA oscillator strengths of plasmon excitations 
of 1D quantum wire and 2D quantum well
electron systems. 
The input parameters are taken from GaAs-based materials:
density $n=10^5{\rm cm}^{-1}$ and lateral width 
$a=b=200\AA$ for the quantum wire; density
$n=1.6\times10^{11}{\rm cm}^{-2}$ 
for a zero thickness purely 2D quantum well.
}
\label{f12}
\end{figure}

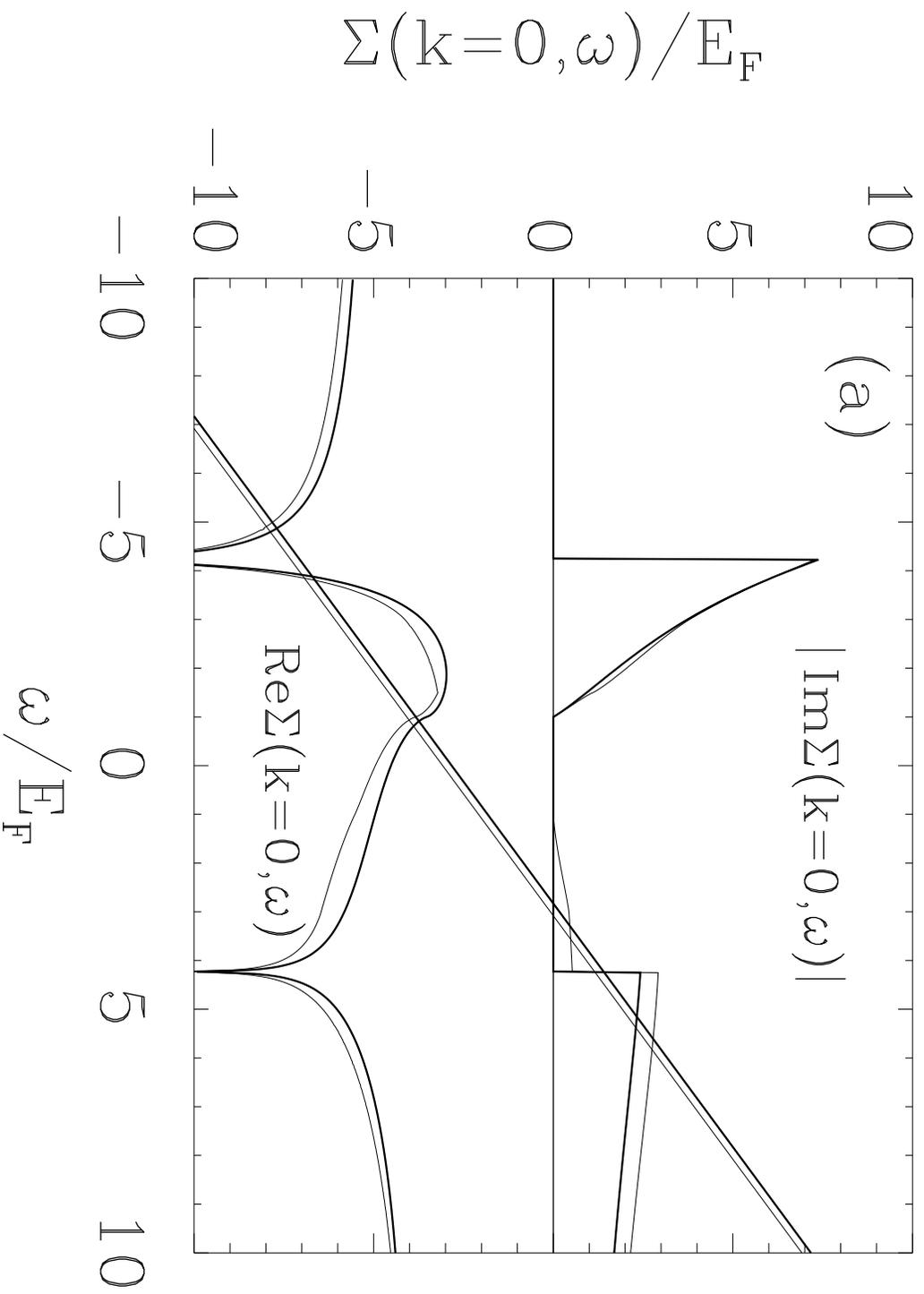
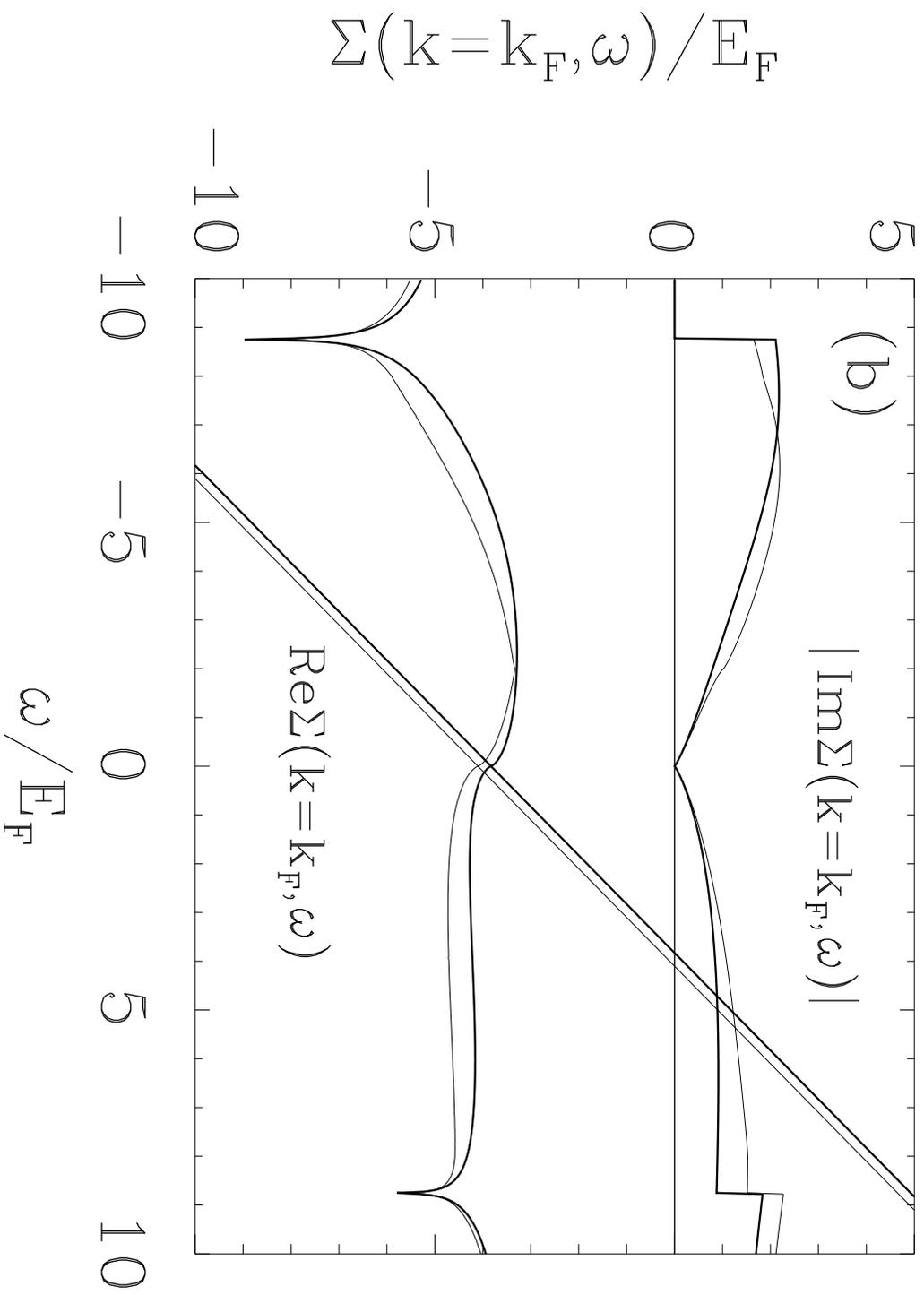
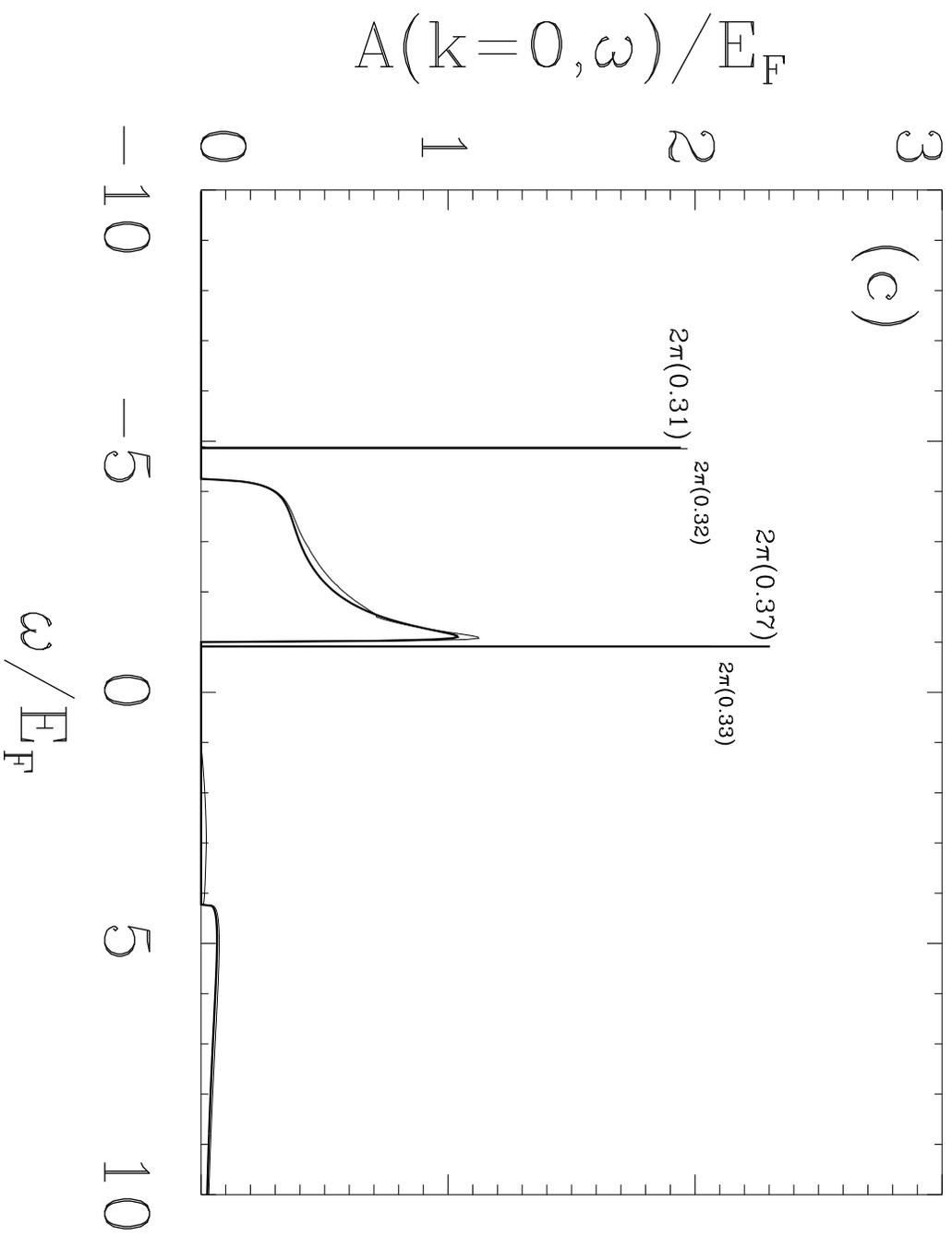
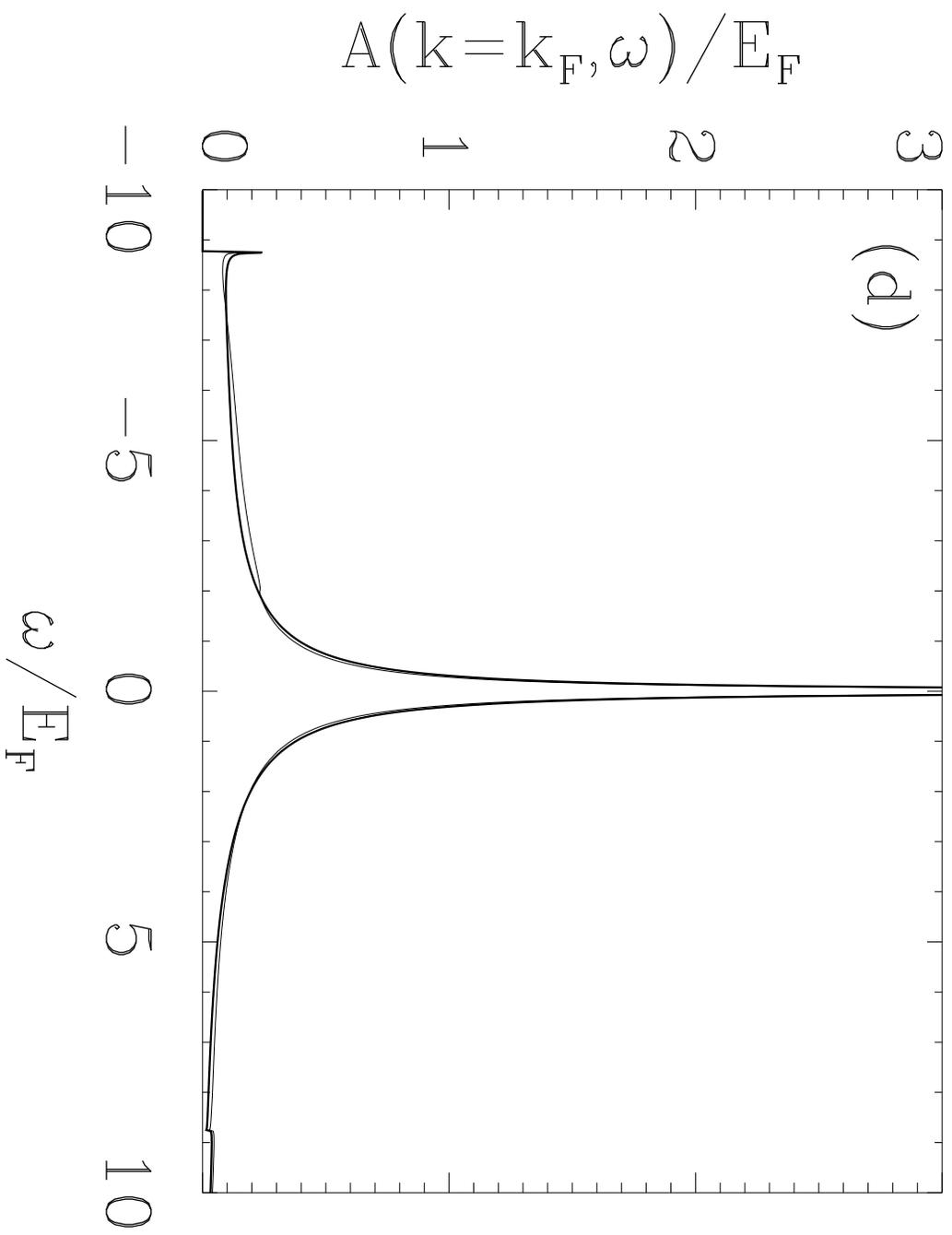
\begin{figure}
\caption{(a), (b) Self-energy $\Sigma(k,\omega)$ and (c), (d) spectral
function $A(k,\omega)$ as functions of the frequency $\omega$ for two
fixed wave vectors $k=0$ ((a) and (c)) and $k_F$ ((b) and (d)). Thick
(thin) lines correspond to the PPA (RPA) results. The vertical lines
in (c) represent $\delta$-functions with 
the spectral weight given above
the peaks. The straight lines in (a) and (b)  are
given by $\omega - \xi(k) 
-\mu$, and their intersections with Re$[\Sigma]$ indicate the solutions
to Dyson's equation and correspond to quasiparticle peaks.
$|{\rm Im}\Sigma|$ is plotted instead of Im$\Sigma$ for visual clarity.}
\end{figure}

\begin{figure}
\caption{
Energy relaxation rates per electron as functions of
electron temperature $T$. The dotted-line,
dot-dashed-line, and solid-line are respectively the results with
no phonon renormalization, 
with phonon renormalization in the plasmon-pole approximation, and 
phonon renormalization in the full RPA.
The quantum wire has a carrier density $n=10^5{\rm cm}^{-1}$
and wire widths $a=b=200\AA$.
}
\label{f31}
\end{figure}

\end{document}